\documentclass[journal]{IEEEtran}
\usepackage{blindtext}
\usepackage{graphicx}
\usepackage{multirow}
\usepackage{latexsym}
\usepackage{amsmath}
\usepackage{amsfonts}
\usepackage{amssymb}
\usepackage{graphicx}
\usepackage{times}
\usepackage[compress]{cite}
\usepackage{color}
\usepackage{subfigure}
\usepackage[numbers]{natbib}
\usepackage[hidelinks]{hyperref}
\usepackage{colortbl}
\usepackage{xcolor,soul}
\setcitestyle{square}

\newtheorem{remark}{Remark}

\newtheorem{lemma}{Lemma}
\newtheorem{thm}{Theorem}
\newtheorem{assumption}{Assumption}

\ifodd 1

\newcommand{\com}[1]{{\color{red}{Comment: #1}}}
\else
\newcommand{\com}[1]{}
\fi

\ifCLASSINFOpdf
\else
\fi

\begin{document}
\title{Reliability and Market Price of Energy in the Presence of Intermittent and Non-Dispatchable Renewable Energies}

\author{Ashkan Zeinalzadeh, Donya Ghavidel, and Vijay Gupta\thanks{The authors are with the Department of Electrical Engineering, University of Notre Dame, IN, USA. Email:  (\texttt{azeinalz,dghavide,vgupta2)@nd.edu.} A preliminary version of these results were accepted at American Control Conference, 2018~\cite{PricingEnergy}. }}

\maketitle

\begin{abstract}
The intermittent nature of the renewable energies increases the operation costs of conventional generators. As the share of energy supplied by renewable sources increases, these costs also increase. In this paper, we quantify these costs by developing a market clearing price of energy in the presence of renewable energy and congestion constraints. We consider an electricity market where generators propose their asking price per unit of energy to an independent system operator (ISO). The ISO solve an optimization problem to dispatch energy from each generator to minimize the total cost of energy purchased on behalf of the consumers. 

To ensure that the generators are able to meet the load within a desired confidence level, we incorporate the notion of load variance using the Conditional Value-at-Risk (CVAR) measure in an electricity market and we derive the amount of committed power and market clearing price of energy as a function of CVAR. 
It is shown that a higher penetration of renewable energies may increase the committed power, market clearing price of energy and consumer cost of energy due to renewable generation uncertainties. We also obtain an upper-bound on the amount that congestion constraints can affect the committed power. We present  descriptive simulations to illustrate the impact of renewable energy penetration and reliability levels on committed power by the non-renewable generators, difference between the dispatched and committed power, market price of energy and profit of renewable and non-renewable generators.
\end{abstract}


\IEEEpeerreviewmaketitle

\section{Introduction}
Growing concern over the effects of climate change has caused a noticeable shift from non-renewable resources in many parts of the world. As a result, renewable energy resources are anticipated to play a crucial role in power systems of the near-future. However, integration of renewable resources into the electricity market, specifically high levels of penetration, requires several modifications in the electricity market \cite{kirschen}. The primary reason for this is that a significant portion of the load will be met by renewables with zero marginal costs with higher penetration of renewable energies. The production from  renewable energy sources is highly uncertain and variable and will eventually necessitate a higher capacity of reserves to ensure grid reliability. Therefore, a higher penetration of renewable energy not only decreases the revenue of the non-renewables but also can increase the maintenance and operating cost of the non-renewable generators in a long time-horizon. Thus, the electricity market should be modified to take this uncertainty and variability into account and to mitigate the impact on other entities involved in the market. Despite available methods for mitigating uncertainty, e.g. energy storage systems, demand response etc., the uncertainties of renewable generation remain as challenges to optimizing generator dispatch in the day-ahead market (see \cite{jones2017, graf} and the references therein). %Energies generated by wind and solar have higher and more rapid fluctuations than the load. Therefore, unlike the load, forecasting of distributed renewable generation in a large scale grid remains as a challenge for the dispatch operators. 

In this paper, we consider an electric grid that delivers demanded electricity to consumers. The grid consists of non-renewable generator units, renewable generator units, and consumers. The non-renewable generator units are composed of a set of generators with different ramp constraints and cost parameters. The non-renewable and renewable generators are operated by different agents. We assume that the grid takes all generated renewable energy, unless the total renewable generation is more than the aggregate load. Renewable generations in excess of the load are curtailed. The non-renewable generator is dispatched to meet the net-load, which is the consumer load minus the renewable generator output. We also assume that the load is not price sensitive and that load service entities submit their hourly demand schedules to the ISO.

This work proposes an analytical method to study the effect of (non-dispatchable) renewable generation capacity in the dispatch strategy while considering grid reliability. For the sake of simplicity and to enable the development of an analytical solution, we consider a radial grid. In this paper, we consider renewables with zero marginal cost and we assume renewables are paid based on the market price of energy. In our model, renewable energy production is described by a stochastic process. The main contribution of this work is to quantify the required power and market price of energy as a function of renewables and loads statistics. The most important result of this study is that the market clearing price of energy is shown to be a non-decreasing function of the uncertainty of the net load, and the reliability demanded by the consumer. We derive an analytical solution for the amount of committed power to meet the load within a desired level of reliability. This analytical solution describes how the committed power is affected by the tail distribution of the net-load. Higher penetration of renewable energy increases the tail distribution of the net-load subsequently, the total committed power increases. Additionally, the combination of uncertainties in renewables generation and congestion constraints increases the committed power for non-renewables. We derive an analytical upper-bound on the committed power resulting from net-load uncertainties and congestion constraint. To illustrate the ideas, we first  study how the market price of energy, generators profit and committed power varies for different desired level of reliability at a fixed penetration level of the renewables. We observe that committed power, and price are increasing in the desired level of reliability. A higher reliability level leads to a higher planning for the committed power and increases the expected deviation from committed power, as well as associated cost. We denote this cost as the 'deviation cost' and obtain this cost as a function of the market price of energy. We are interested to see how this deviation cost varies as the penetration of renewables increases. Then, we study how the market price of energy, generator profits and committed power vary for different levels of penetration for a fixed reliability. The total amount of committed power of non-renewable energy first decreases as the penetration of the renewables increases because the net-load decreases. After reaching a threshold, the required power from non-renewables increases due to uncertainty about the production of renewables. It is also shown that the deviation cost is increasing in the penetration level. This observation may be counter-intuitive as the preliminary expectation is that the market price of energy decreases as the penetration of the renewables increases, because of the lower marginal cost of renewables. However, the uncertainties and intermittencies of renewables are increasing in the penetration level which in turn can increase the cost.

The rest of the paper is structured as follows. In Section~\ref{sec:relunc}, we formulate and solve the optimization problem of the ISO with the goal of minimizing the cost of energy while ensuring the planned generators are able to meet the load within a desired confidence level. In Section~\ref{sec:cong}, we study the effect of congestion on the electricity market and obtain an upper bound on the risk-based committed power. In Section~\ref{sec:nonren}, we formulate the profit function of the generators and describe the reasons of decreasing the profit of the generators at a high level of the renewables penetration. In Section~\ref{sec:IV}, we illustrate numerically the effect of the renewable penetration and reliability levels on the market price of energy, generators' profit and deviation cost and the recovery costs. Finally, concluding remarks are provided in Section~\ref{sec:V}.% and supplementary proofs are provided in the appendix, Section~\ref{sec:VI}. 

%Consequently, if the consumers impose the constraint of a maximum price they are willing to pay for energy, a monetary transfer from the renewable producer to the non-renewable producer may be required to make the non-renewable producer whole. This transfer characterizes the externality that the renewable producer imposes on the grid and is absorbed by the non-renewable producer. If the price paid by the consumer is especially low, this may be a natural cap on the level of renewable penetration. 

%Although the model we consider is stylized, the conclusions are striking and may point to a fundamental rethinking of the way renewable portfolio standards are currently planned.

%To illustrate the ideas, we provide numerical examples in Section~\ref{sec:IV}. In Section~\ref{sec:IV}, we analyze the impact of renewable energy penetration on the committed power as well as the expected dispatched power of both the non-renewable generators and renewable generators.  The deviation from committed power and the expected dispatched power by the non-renewables is observed to be increasing in renewables penetration. This deviation may increase the reserve cost, and subsequently increases the price of energy. 

\subsection{Related literature}
 There exists a rich literature which studies various aspects of renewable energy resources in the power grid that includes forecasting methods, energy storage, frequency regulation and technological challenges
(see, e.g.,~\cite{wang, carrasco, bevrani} and the references therein). Further, there is also a trend of papers which focus on the integration of renewable energy producers into electricity markets, their competition and the their impact on the market (see \cite{klessmann, menanteau, ramesh}). Of particular relevance, the work in \cite{klessmann} investigates the integration of renewable energy resources, specifically wind power in Germany, Spain, and the UK.
 The authors in \cite{menanteau} focuse on the efficiency of incentive schemes for the development of renewable energy sources and their integration into the electricity market. The authors in \cite{ramesh} study the strategic behaviors of renewable power producers in electricity markets.

 There exists another line of related literature which answers the question of how to allocate the cost generated by the uncertainty/variability of the renewable energy producers or the benefit produced by their aggregation among them, satisfying certain properties in the electricity market (see e.g.,~\cite{baeyens2011,baeyens2013, lin2014}). Authors in \cite{lin2014} propose an ex post cost sharing scheme which allocates aggregate system cost back to variable energy producers, proportional to their marginal contribution to the aggregate system imbalance, while satisfying certain axioms in the forward market for energy. 

 In this paper, we use the notion of Conditional Value-at-Risk (CVAR) to quantify the effect of the uncertainty of the net-load on market price of energy. CVAR has been used in the electricity market and literature (see, e.g. \cite{Bukhsh}) to measure the risks of dispatch strategies. Several works that are representative of the direction of this study include \cite{Bitar, Bitar1, bathurst, Bilateral, PricingEnergy,naturalgas}, %Risk, %Joskow,morales,botterud,matevosyan}, 
 which analyze the problem under different market settings. The studies most relevant to our work are \cite{Bitar} and \cite{Bitar1}. They develop optimal strategies to inject wind energy into the grid under a fixed market price of energy. Unlike these works, we develop the market clearing price of energy to quantify the effect of uncertainty of the net-load on the market clearing price of energy. 

%\rev{our main contributions:\begin{itemize}\end{itemize}}

\subsection{Notation}
The following notations are used in this work. 

\begin{itemize}
\item $N:$ Number of buses,
%\item $M_{i}:$ Number of non-renewable generators at bus $i$,
\item $w_{i}:$ Capacity of renewable energy at node $i$,
\item $W=(w_1,...,w_{N})$,
\item $K_{W}:$ Number of contingency scenarios,
%\item $\theta_{i}^{tk}:$ Voltage phase at bus $i$, time $t$ and scenario $k$,
%\item $v_{i}^{tk}:$ Voltage amplitude at bus $i$, time $t$ and scenario $k$,
\item $C_{i}:$ Set of all connected buses to bus $i$,
\item $p_{d_{i}}^{t}:$ Active load at bus $i$ and time $t$,
%\item $q_{d_{i}}^{tk}:$ Reactive load at bus $i$, time $t$ and scenario $k$,
%\item $p_{g_{i}}^{t}:$ Total non-renewable power generated at bus $i$,
\item $p_{r_i}^{t}:$ Active power generated by renewables at bus $i$ and time $t$,
%\item $q_{r_i}^{tk}:$ Reactive power generated by renewables at bus $i$, time $t$ and scenario $k$,
\item $p_{g_{i}}^{t}:$ Active power generated by non-renewable generator at bus $i$ and time $t$, %we abbreviate $p_{g_{i}}^{tk}$ with $p_{g_{i}}^{t}$ if the number of contingency scenarios is one ($K_{W}=1$),
\item $\pi_{i}:$ Asking price per unit of energy of the generating unit at bus $i$, 
\item $\pi_{r}$: Asking price per unit of renewable energy,
\item $p_{ij}^{t}:$ Active power flow from the line connecting the bus $i$ to $j$ at time $t$,
\item $S_{i}:$ The generator start-up cost at bus $i$, 
\item $\beta_{i}:$ The no-load cost per hour for generator at bus $i$,
\item $b_{ij}:$ Admittance of the line connecting bus $i$ to $j$,
\item $p_{g_{i}}^{\max}:$ Maximum power of generator at bus $i$,
\item $p_{g_{i}}^{\min}:$ Minimum power of generator at bus $i$,
%\item $p_{i}^{tk}:$ Active power at bus $i$, time $t$ and scenario $k$,
\item ${p}_{ij}^{\max}:$ Maximum active power loss for the line connecting bus $i$ to $j$, 
\item $rp_{g_{i}}^{t}:$ Reserve of generator at bus $i$ and time $t$,
%\item $rp_{g_{i}}^{-t}:$ Reserve down of generator at bus $i$ and time $t$,
\item $rp_{g_{i}}^{\max}:$ Maximum reserve of generator at bus $i$,
%\item $rp_{g_{i}}^{-\max}:$ Maximum reserve down of generator at bus $i$,
\item $\delta p_{g_{i}}^{t}:$ Ramp up/down of generator at bus $i$ and time $t$,
%\item $\delta p_{g_{i}}^{-t}:$ Ramp down of generator at bus $i$ and time $t$,
\item $\delta p_{g_{i}}^{\max}:$ Maximum ramp up/down of generator at bus $i$,
%\item $\delta p_{g_{i}}^{-\max}:$ Maximum ramp down of generator at bus $i$,
%\item $p_{g_{i}}^{tc}:$ Power contract with generator at bus $i$ and time $t$,
%\item $\Delta p_{g_{i}}^{tk}:$ Deviation from power contract $p_{g_{i}}^{tc}$ of generator at bus $i$, time $t$ and scenario $k$,
\item $(.)^+=\max(.,0)$
\item $s_{i}^{t}:=p_{d_{i}}^{t}-p_{r_{i}}^{t}$,
%\item $Z=\big{\{} (i,t,k) \big{|} 1 \leq i \leq N, 1 \leq t \leq  T, 1 \leq k \leq K \big{\}}$,
%\item $\Theta=\{\theta_{i}^{tk} \big{|} (i,t,k) \in Z\}$,
%\item $P_{D}=\{p_{d_i}^{tk} \big{|} (i,t,k) \in Z\}$,
%\item $P_{r}=\{p_{r_i}^{tk} \big{|} (i,t,k) \in Z\}$,
%\item $\tilde{Z}=\big{\{} (i,t,k) \big{|} 1 \leq i \leq N, 1 \leq t \leq  T, 1 \leq k \leq K \big{\}}$, 
\item $P=\{ p_{g_{i}}^{t} \big{|} 1 \leq i \leq N, 1 \leq t \leq  T \}$, 
%\item $\bar{Z}=\big{\{} (i,t) \big{|} 1 \leq i \leq N, 1 \leq t \leq  T \big{\}}$,
%\item $P^{c}=\{ p_{g_{i}}^{tc} \big{|} (i,t) \in \bar{Z}, \text{ where the generator } i\\ \text{ can be either dispatchable  or non-dispatchable}\}$,  
\item $I_{i}^{t}:$ An indicator of the event that the generator at bus $i$ is online at time $t$,
%\item $I=\{I_{i}^{t} \big{|} (i,t) \in \bar{Z}\}$, 
%\item $I_{i}^{t}=(I_{i1}^{t},...,I_{iM_{i}}^{t})$, 
%\item $\delta P=\{ \delta p_{g_{i}}^{t} \big{|} (i,t) \in \bar{Z}\}$, 
%\item $RP=\{ rp_{g_{i}}^{t} \big{|} (i,t) \in \bar{Z}\}$,
%\item $P_{i}^{t}=(p_{g_{i1}}^{t},...,p_{g_{iM_{i}}}^{t})$, 
%\item $P^t=(p_{g_{1}}^{t},...,p_{g_{N}}^{t})$,
%\item $T^{u_{ij}}:$ Minimum up time for the $j$th generator (at bus $i$), 
%\item $T^{d_{ij}}:$ Minimum down time for the $j$th generator (at bus $i$), 
%\item $u_{ij}^{t}:$ Number of stages that the $j$th generator (at bus $i$) has been online from its most recent turning on to the beginning of stage $t$,
%\item $d_{ij}^{t}:$ Number of stages that the $j$th generator (at bus $i$) has been off, from the most recent turning off to the beginning of stage $t$,
%\item If there is no contingency scenario ($K_{W}=1$) we abbreviate $p_{i}^{tk}$, $\theta_{i}^{tk}$, $v_{i}^{tk}$, $p_{r_i}^{tk}$, $p_{d_i}^{tk}$, $p_{g_{i}}^{tk}$ and $p_{ij}^{tk}$ with $p_{i}^{t}$, $\theta_{i}^{t}$, $v_{i}^{t}$, $p_{r_i}^{t}$, $p_{d_i}^{t}$, $p_{g_{i}}^{t}$ and $p_{ij}^{t}$ respectively.
\item $P^{K}=\{ p_{g_{i}}^{tk} \big{|} 1 \leq i \leq N, 1 \leq t \leq  T, 1 \leq k \leq K_{W} \}$. 
\end{itemize}

\section{Reliability and Renewable Uncertainty} \label{sec:relunc}

In this section, we formulate the dispatch optimization problem for a case where the renewable energy resource is present in the electricity market. We first consider the deterministic load and renewable generator then extend it to the stochastic model for load and renewable energy generation. The ISO dispatches the generators in the market based on the marginal costs. %We approximate the term in the bracket in (\ref{eq:dispatchopt}) with marginal cost $\pi_{ij}$. 
We first assume that the load $p_{d_i}^{t}$ and renewable energy generation $p_{r_i}^{t}$ are deterministic and there is no contingency scenario. We then formulate the dispatch optimization problem and extend it to a scenario in which a stochastic model is considered for the load and renewable energies. To develop an analytical solution we consider Assumption~\ref{assumption1}.

\begin{assumption}
\label{assumption1}
\,\  
\begin{itemize}
%there is no contingency scenario ($K_{W}=1$)
%\item a) we consider a continuous model for load and renewable realizations and abbreviate $\theta_{i}^{tk}$, $v_{i}^{tk}$, $p_{r_i}^{tk}$, $p_{d_i}^{tk}$, $p_{g_{i}}^{tk}$ and $p_{ij}^{tk}$ with $\theta_{i}^{t}$, $v_{i}^{t}$, $p_{r_i}^{t}$, $p_{d_i}^{t}$, $p_{g_{i}}^{t}$ and $p_{ij}^{t}$ respectively.
\item a) $ 0 \leq \pi_{r} < \pi_{1} < ...<\pi_{N}$.
%\item b) we let $p_{g_{i}}^{t}$ to be the committed power of generator at node $i$. $p_{g_{i}}^{tc}=p_{g_{i}}^{t}$.
\item b) $\sin(\theta_i^t-\theta_j^t)\approx \theta_i^t-\theta_j^t$, and $v_i^t \approx 1$.  
%\item e) the reactive power load and generation are zero $q_{r_i}^{tk}=q_{d_i}^{tk}=0$.
\end{itemize}
\end{assumption} 
\subsection{Deterministic Load and Renewable Generation}
%If Assumption \ref{assumption2} holds, then the constraints (\ref{ineq:gencons})-(\ref{ineq:activecont}) are reduced to 
%(\ref{constnum1})-(\ref{ineq:cons}):
The output of the $i$th generator satisfies the following constraints
\begin{align}\label{constnum1}
p_{g_{i}}^{\min} \leq p_{g_{i}}^t \leq p_{g_{i}}^{\max}. 
\end{align}
If part $b)$ of Assumption~\ref{assumption1} holds then the active power injected (or withdrawn) from bus $i$ at time $t$ is given as
\begin{align}
\label{eq:balance} 
&p_{g_{i}}^{t}+p_{r_i}^{t}-p_{d_i}^{t}=\sum_{j \in C_{i}} b_{ij}  (\theta_i^t-\theta_j^t),
\end{align}
The active power flow $p_{ij}^{t}$ for the line connecting bus $i$ to bus $j$ at time $t$ must satisfies
\begin{align}
\label{ineq:cons}
p_{ij}^{t}=b_{ij} (\theta_{i}^{t}-\theta_{j}^{t}) \leq {p}_{ij}^{\max}.
\end{align}
The ISO minimizes the total cost of energy on behalf of the consumers as follows  
\begin{align}\label{optim}
\mathcal{P}_{1}: \underset{w.r.t. \,\ (\ref{constnum1})-(\ref{ineq:cons})}{\underset{P}{\min}}  \,\ \sum_{t=1}^{T} \sum_{i=1}^{N} \big{[} \pi_{i} p_{g_{i}}^{t}+\pi_{r} p_{r_i}^{t} \big{]}.
\end{align}
The solution to problem $\mathcal{P}_{1}$ is provided in Theorem~\ref{thm1}.
\begin{thm}
\label{thm1}
Let $(\mu_{i}^t,\bar{\mu}_{i}^t)$, $\lambda_i^t$, and $\tilde{\mu}_{ij}^{t}$ be the Lagrange multipliers corresponding to (\ref{constnum1}), (\ref{eq:balance}), and (\ref{ineq:cons}) respectively. For deterministic load and renewable generation, the solution to problem $\mathcal{P}_{1}$ is given by 
\begin{align*}
& \pi_{i}-\lambda_i^t+\mu_{i}^{t}-{\bar{\mu}}_{i}^{t}=0, \\ \notag
& \sum_{j \in C_i} b_{ij} [ \tilde{\mu}_{ij}^t-{\tilde{\mu}}_{ji}^{t}+\lambda_i^t-\lambda_j^t]=0, \\ \notag
&p_{r_i}^{t}-p_{d_i}^{t}+ p_{g_{i}}^{t}=\sum_{j \in C_{i}} b_{ij} (\theta_i^t-\theta_j^t),\\
&b_{ij} (\theta_i^t-\theta_j^t) \leq {p}_{ij}^{\max}, \quad \tilde{\mu}_{ij}^{t} \big{(}b_{ij} (\theta_i^t-\theta_j^t) - {p}_{ij}^{\max}\big{)}=0, \\
& p_{g_{i}}^{\min} \leq p_{g_{i}}^{t} \leq p_{g_{i}}^{\max}, \quad \mu_{i}^{t} (p_{g_{i}}^{t}-p_{g_{i}}^{\max})=0,  \\ \notag 
& \bar{\mu}_{i}^{t}(p_{g_{i}}^{\min}-p_{g_{i}}^{t})=0, \quad \mu_{i}^{t} \geq 0, \quad \bar{\mu}_{i}^{t} \geq 0,
\end{align*}
where $(\mu_{i}^t,\bar{\mu}_{i}^t)$, $\lambda_i^t$, and $\tilde{\mu}_{ij}^{t}$ represent the Lagrange multipliers corresponding to (\ref{constnum1}), (\ref{eq:balance}), and (\ref{ineq:cons}) respectively.
\end{thm}
\begin{remark}
Note that $\lambda_{i}^t$ represents the locational marginal price at bus $i$ and time $t$.
\end{remark}
\begin{proof}
See Appendix.
\end{proof}

 In the following section, we describe analytically how the uncertainties in the renewable energy generation change LMPs. Below, we extend the optimization problem $\mathcal{P}_{1}$ for a scenario in  which the loads and renewable generations are described as continuous stochastic processes.

\subsection{Stochastic Model for Load and Renewable Energy} \label{sec:stochasticmodel}

Let $p_{g_{i}}^{\min}=0$. Without the congestion constraint (\ref{ineq:cons}), ${p}_{ij}^{\max}=\infty$, all locational marginal prices are equal to the market clearing price of energy $\lambda_i^t=\lambda_j^t=\lambda^t$. Therefore, the power balance equation (\ref{eq:balance}) for all $i=1,...,N$ is reduced to
\begin{align}\label{wtcon}
\sum_{i=1}^{N} p_{g_{i}}^{t}=\sum_{i=1}^{N} (p_{d_i}^{t}-p_{r_i}^{t}).
\end{align}

Now, we extend the power balance equation~(\ref{eq:balance}) to a scenario in which load $p_{d_i}^{t}$ and renewables $p_{r_i}^{t}$ have continuous stochastic models. Let $\{(p_{d_{1}}^{t},...,p_{d_{N}}^{t})\}_{t=1}^{T}$, and $\{(p_{r_{1}}^{t},...,p_{r_{N}}^{t})\}_{t=1}^{T}$ be non-negative random variables with known continuous probability density functions on the probability space $(\Omega, \mathfrak{F}, \mathbb{P})$. For the given $o \in \Omega$, $\{(p_{d_{1}}^{t}(o),...,p_{d_{N}}^{t}(o))\}_{t=1}^{T}$, and $\{(p_{r_{1}}^{t}(o),...,p_{r_{N}}^{t}(o))\}_{t=1}^{T}$ are deterministic functions of $t$ that denote the load realization and renewable resource power realization respectively. The committed generators in the market must be able to meet the power balance (\ref{eq:stbalance}) within the desired level of reliability $\alpha \in [0 \,\ 1]$. This means that sufficient resources must be planned in the scheduling problem such that load can be met for a possible subset of realizations $\tilde{\Omega} = \{ o | o \in \Omega\}$, where the probability of $\tilde{\Omega}$ is at least $\alpha$. These resources ensure that for any given $o \in \tilde{\Omega}$ there exists a solution  $p_{g_{ij}}^{t}$ for the dispatch optimization problem such that (\ref{eq:stbalance}) holds:
\begin{align}\label{eq:stbalance} 
p_{g_{i}}^{t}=p_{d_i}^{t}(o)-p_{r_i}^{t}(o)+\sum_{j \in C_{i}} Y_{ij}  (\theta_i^t-\theta_j^t),
\end{align}
for all $i=1,...,N$ and $t=1,...,T$. Below, we reformulate the dispatch optimization problem with the new power balance constraint (\ref{eq:stbalance}). 

Let $s_{i}^{t}= p_{d_i}^{t}-p_{r_i}^{t}$ and $n^{t}=\big{(}  \sum_{i=1}^{N} (s_i^{t}- p_{g_{i}}^{t}) \big{)}^{+}$. Define $F_{n^t}^{W}$ as the cumulative distribution functions of $n^t$, that depends on the capacity of renewables $W$. We use the concept of Value-at-Risk (VaR) and Conditional Value-at-Risk (CVaR) \cite{Rockafellar} to obtain the upper bounds on the contingency scenarios. $VaR_{\alpha}^{W}(n^t)$ determines the worst possible $n^t$ that may occur within a given confidence level $\alpha$.
%\begin{definition}
For a given $0 < \alpha < 1$, the amount of $n^t$ will not exceed $VaR_{\alpha}^{W}(n^t)$ with probability $\alpha$, $VaR_{\alpha}^{W}(n^t)=\min \{z| F_{n^t}^{W}(z) \geq \alpha \}$. $CVaR_{\alpha}^{W}$ is defined as the conditional expectation of $n^t$ above the amount $VaR_\alpha^{W}$. Let $E$ denote the expectation over $n^t$, $CVaR_{\alpha}^{W}(n^t)=E [ n^t | n^t > VaR_{\alpha}^{W}(n^t) ]$, and $
CVaR_{\alpha}^{W}(n^t)=\int_{-\infty}^{\infty}z dF_{n^t}^{\alpha}(z)$, where
\[
    F_{n^t}^{W,\alpha}(z)=
\begin{cases}
    0,& \text{if }  z < VaR_{\alpha}^{W}(n^t) \\
    \frac{F_{n^t}(z)-\alpha}{1-\alpha}, & \text{otherwise}
\end{cases}.
\]
%\end{definition}
In our dispatch optimization problem the objective is to plan for the generators such that they are capable of meeting the load within the confidence level. We write the condition (\ref{eq:stbalance}) as $CVaR_{\alpha}^{W}(n^t)=0$.

\begin{thm}
\label{thm2}It is claimed that
\begin{align} \label{CVAR}
&CVaR_{\alpha}^{W}(n^{t})=\\ \notag
& \,\ \,\ \,\ \,\ \,\ \,\ \begin{cases}
    0,& \text{if }  n^t \equiv 0 \\
    CVaR_{\alpha}^{W}(\sum_{i=1}^{N} s_i^{t}) - \sum_{i=1}^{N} p_{g_{i}}^{t}, & \text{otherwise}
\end{cases}.
\end{align}
\end{thm}
Proof. The proof is given in the Appendix.

We replace the power balance equation (\ref{eq:stbalance}) with
\begin{align}\label{eq:balancee}
&CVaR_{\alpha}^{W}(n^{t})=0,\\ \label{eq:balanceee}
&\sum_{i=1}^{N} p_{g_{i}}^{t}=CVaR_{\alpha}^{W}(\sum_{i=1}^{N} s_i^{t}).
\end{align}
The implication of equation (\ref{eq:balanceee}) is that if the total planned power is equal to $CVaR_{\alpha}^{W}(\sum_{i=1}^{N} s_i^{t})$ then the load will be met with confidence level $\alpha$.

%Based on the PJM guidelines (\cite{PJM2}-\cite{PJM}), generators that are located at one bus are considered as one unit and must offer one set of data including start-up, no-load and marginal cost to the PJM. Therefore, we define the minimum and maximum generated power at bus $i$ as follows:
%\begin{align}
%& p_{g_{i}}^{\min}=\min_{1 \leq j \leq M_{i}} p_{g_{ij}}^{\min},\\
%& p_{g_{i}}^{\max}=\sum_{j=1}^{M_{i}} p_{g_{ij}}^{\max}.
%\end{align}
%We replace the generator constraint (\ref{constnum1}) with
%\begin{align}\label{constnum2}
%p_{g_{i}}^{\min} \leq p_{g_{i}}^t \leq p_{g_{i}}^{\max}.
%\end{align}
%We define the marginal cost of the generator at bus $i$ as 
%\begin{align}
%\pi_i= \frac{\sum_{j=1}^{M_{i}} (p_{g_{ij}}^{\max}-p_{g_{ij}}^{\min})\pi_{{ij}}}{\sum_{j=1}^{M_{i}} (p_{g_{ij}}^{\max}-p_{g_{ij}}^{\min})}.
%\end{align}
%It can be concluded from part $b)$ of Assumption~$2$ that
%\begin{align}
%0 \leq \pi_{r} < \pi_{1} < \pi_{2} < ... < \pi_{N}.
%\end{align}
The ISO minimizes the total cost of energy on behalf of the consumers as follows
\begin{align}\label{optimm}
\mathcal{P}_{2}: \underset{w.r.t. \,\ (\ref{constnum1}), (\ref{eq:balanceee})}{\underset{(P^{1},...,P^{T})}{\min}} \underset{P_{D},P_{r}}{E} \,\ \sum_{t=1}^{T} \sum_{i=1}^{N} \big{[} \pi_{i} p_{g_{i}}^{t}+\pi_{r_i} p_{r_i}^{t} \big{]}.
\end{align}

We first ignore the congestion constraint (\ref{ineq:cons}) and solve the optimization problem (\ref{optimm}). This assumption is reasonable if the economical profit of relaxing the congestion privilege on its cost. We will study the effect of congestion constraint on the planned power in the Section~\ref{sec:cong}.

To ensure a feasible solution for the ISO's problems (\ref{optimm}), Assumption~\ref{assumption2} is considered. It is worth noting that this assumption must be modified for a more complex topology of the power grid, e.g. with congestion constraints.

\begin{assumption}
\label{assumption2}
We assume that
\begin{itemize}
\item[a)] For all $t=1,...,T$
\begin{equation}
\underset {1 \leq i \leq N}{\min} \,\ p_{g_{i}}^{\min} \leq CVaR_{\alpha}^{W}(\sum_{i=1}^{N} s_{i}^t) \leq \sum_{i=1}^{N} p_{g_i}^{\max}.    
\end{equation}
\item[b)] $\underset{i \in \{1,...,N\}}{\max} p_{g_i}^{\min} < \underset{{i \in \{1,...,N\}}}{\min} \{ p_{g_i}^{\max}-p_{g_i}^{\min}  \}$.\\
\end{itemize}
\end{assumption} 

Because of part $a)$ of Assumption~\ref{assumption2}, there exists an unique $1 \leq k \leq N$ such that
$\sum_{i=1}^{k-1} p_{g_i}^{\max}  < CVaR_{\alpha}^{W}(\sum_{i=1}^{N} s_{i}^t) \leq \sum_{i=1}^{k} p_{g_i}^{\max}$. Part $b)$ of Assumption~\ref{assumption2} ensures that if the first $(k-1)$th generators are operating at their maximum power and additional power is needed to meet the load, when it is less than $p_{g_k}^{\min}$, then generator $(k-1)$th
can lower its output power without violating its output power constraint; such that the $k$th generator operates at its minimum power ($p_{g_{k}}^{\min}$).
This is proved in Lemma~$1$, in below, and is drawn from part $b)$ of Assumption~\ref{assumption2}

\begin{lemma} 
\label{lemma1}If $0 < CVaR_{\alpha}^{W}(\sum_{i=1}^{N} s_{i}^t)-\sum_{i=1}^{k-1} p_{g_{i}}^{\max} < p_{g_k}^{\min}$ then
\begin{equation}\label{neq:lem1}
p_{g_{k-1}}^{\min} < CVaR_{\alpha}^{W}(\sum_{i=1}^{N} s_{i}^t)-\sum_{i=1}^{k-2}p_{g_i}^{\max}-p_{g_{k}}^{\min} < p_{g_{k-1}}^{\max}.\\
\end{equation}
\end{lemma}
Proof. The proof is given in the Appendix.

\begin{thm}\label{thm3}
Let $(\mu_{i}^t,\bar{\mu}_{i}^t)$ and $\lambda^t$ be the Lagrange multipliers corresponding to (\ref{constnum1}) and (\ref{eq:balanceee}) respectively. If Assumptions~ \ref{assumption1} and \ref{assumption2} hold then the solution to $\mathcal{P}_{2}$ is given by

\begin{itemize}
\item If $p_{g_k}^{\min} \leq CVaR_{\alpha}^{W}(\sum_{i=1}^{N} s_{i}^t)-\sum_{i=1}^{k-1} p_{g_{i}}^{\max} \leq p_{g_k}^{\max}$
\begin{align}
&\mu_i^t >0, \bar{\mu}_i^t=0, p_{g_i}^t=p_{g_i}^{\max} \text{ for  \,\ } i=1,...,k-1,\\
&\mu_k^t=0, \bar{\mu}_k^t=0, p_{g_k}^t=CVaR_{\alpha}^{W}(\sum_{i=1}^{N} s_{i}^t)-\sum_{i=1}^{k-1} p_{g_{i}}^{\max}, \\
&\mu_i^t=0, \bar{\mu}_i^t=0, p_{g_i}^t=0, \text{ for all } i=k+1,...,N,\label{eq:SOKKT5}\\
&\lambda^t=\pi_i+\mu_i^t-\bar{\mu}_i^t, \,\ \text{for all} \,\ i=1,...,k, \\ \label{price1}
&\lambda^t={\pi_k}. 
\end{align}

\item  If $ 0 < CVaR_{\alpha}^{W}(\sum_{i=1}^{N} s_{i}^t)-\sum_{i=1}^{k-1} p_{g_{i}}^{\max} < p_{g_k}^{\min}$

\begin{align}
&\mu_i^t >0, \bar{\mu}_i^t=0, p_{g_i}^t=p_{g_i}^{\max} \,\ \text{for}\,\ i=1,...,k-2,\\
& \mu_{k-1}^t =0, \bar{\mu}_{k-1}^t=0, \\ \notag
& p_{g_{k-1}}^t=CVaR_{\alpha}^{W}(\sum_{i=1}^{N} s_{i}^t)-\sum_{i=1}^{k-2}p_{g_i}^{\max}-p_{g_k}^{\min}, \\
&\mu_k^t=0, \bar{\mu}_k^t>0, p_{g_k}^t=p_{g_k}^{\min}, \quad \mu_i^t=0, \bar{\mu}_i^t=0, \\
&p_{g_i}^t=0, \text{ for all } i=k+1,...,N,\label{eq:SOKKT5}\\
&\lambda^t=\pi_i+\mu_i^t-\bar{\mu}_i^t, \,\ \text{for all} \,\ i=1,...,k, \\ 
& \lambda^t=\pi_{k-1}.  
\label{eq:price2}
%\label{eq:result}
\end{align}

\item If $0< CVaR_{\alpha}^{W}(\sum_{i=1}^{N} s_{i}^t) <p_{g_1}^{\min}$

From part b) of Assumption~\ref{assumption3}, there exists an $1 \leq k \leq N$ such that 
$p_{g_{k}}^{\min} \leq CVaR_{\alpha}^{W}(\sum_{i=1}^{N} s_{i}^t) <p_{g_{k}}^{\max}$. Let $\bar{k}$ be the smallest $k$ that satisfies this condition, then
\begin{align}
&\mu_i^t=0, \bar{\mu}_i^t=0, p_{g_i}^t=0, \text{ for all } i \neq \bar{k},\\
&\mu_{\bar{k}}^t=0, \bar{\mu}_{\bar{k}}^t=0, p_{g_{\bar{k}}}^t=CVaR_{\alpha}^{W}(\sum_{i=1}^{N} s_{i}^t),  \\ 
&\lambda^t=\pi_{\bar{k}}
\label{eq:price3}.
%\label{eq:resultt}.
\end{align}
\end{itemize}
\end{thm}
\begin{proof}
See Appendix.
\end{proof}

\begin{remark}
It is evident from (\ref{eq:price1}), (\ref{eq:price2}) and (\ref{eq:price3}), that the market clearing price of energy ($\lambda^t$) is higher at times when
$s^t$ has a heavier tail distribution. A heavier tail distribution leads to a higher value of $CVaR_{\alpha}^{W}(\sum_{i=1}^{N} s_{i}^t)$ and larger index of $k$
in (\ref{eq:price1}) and (\ref{eq:price2}). Similarly, a higher level of reliability (larger $\alpha$) leads to a higher market clearing price of energy.
The market clearing prices (\ref{eq:price1}) and (\ref{eq:price2}) are dependent on the tail distribution of $\sum_{i=1}^{N} s_{i}^t$.
The tail distribution of $\sum_{i=1}^{N} s_{i}^t$ depends on the load and renewable energy distributions. Higher penetration of the renewables leads to a heavier tail distribution for the net-load and increase the amount of planned power.% and model of loss function.
\end{remark}

Next, we study the impact of congestion constraint on the planned power for the dispatch problem.

\section{Locational Marginal Prices and Congestion Constraint}\label{sec:cong}

In this section, we study the dispatch optimization problem with the presence of congestion in the power network. We aim to investigate the impact of congestion on the locational marginal prices. Congestion constraints complicate deriving an analytical solution to show the effect of renewables uncertainty on locational marginal prices (LMPs). For the sake of simplicity and to enable us to derive an analytical upper bound on the excess planned power in the current section, we consider a radial grid with $N$ generator buses and consider the following Assumption.
\begin{assumption}: It is assumed that
\label{assumption3}
\,\
\begin{itemize}
\item a) $p_{g_{i}}^{\min}=0$ for all $i=1,...,N$,
%\item b) $0 \leq \pi_{r} < \pi_{1} < ...<\pi_{N}$,
\item b) $CVaR_{\alpha}^{W} (\sum_{k=i}^{N}(p_{d_{k}}^t-p_{r_{k}}^t)) \leq p_{g_{i}}^{\max}$ for all $i=1,...,N$ and $t=1,...T$,
\item c) $p_{ij}^{\max}=\bar{p}$ for all lines.
\end{itemize}
\end{assumption}
Part $b)$ of Assumption~\ref{assumption3} implies that the $i$th generator is not only able to meet its own load but also the aggregate load of the following buses if there are no congestion constraints. This implies that if the congestion constraints were not active then all LMPs are equal to $\pi_{1}^t$.

Let $\hat{p}_{n_i}^t$ be the required non-renewable power to meet the load at bus $i$ within the reliability level $\alpha$ given the dispatch powers $p_{g_{1}}^t,...,p_{g_{i-1}}^t$. Let $\hat{p}_{n_{iN}}^t$ be the required power to meet the loads from bus $i$ to bus $N$ within the reliability level $\alpha$ given the dispatch powers $p_{g_{1}}^t,...,p_{g_{i-1}}^t$. Below, we derive a formula for $p_{g_{i}}^t$ as a function of $\hat{p}_{n_i}^t$ and $\hat{p}_{n_{iN}}^t$.  The powers $\hat{p}_{n_1}^t$ and $\hat{p}_{n_{1N}}^t$ are given as
$\hat{p}_{n_1}^t=CVaR_{\alpha}^{W}(p_{d_1}^t-p_{r_1}^t)$ and $\hat{p}_{n_{1N}}^t=CVaR_{\alpha}^{W}(\sum_{i=1}^{N} (p_{d_i}^t-p_{r_i}^t))$. Based on the values of $\hat{p}_{n_1}^t$ and $\hat{p}_{n_{1N}}^t$ one of the following may hold.

\begin{itemize}
\item Case I: $\hat{p}_{n_{1N}}^t \leq  \hat{p}_{n_1}^t+\bar{p}$%\bar{p}$

If $\hat{p}_{n_{1N}}^t \leq  \hat{p}_{n_1}^t+\bar{p}$ then the generator at bus~$1$ supplies the total demanded power $\hat{p}_{n_{1N}}^t$ without violation of the congestion constraint. Therefore, the remaining generators do not produce any power.
\begin{align}
&p_{g_1}^t=\hat{p}_{n_{1N}}^t, 
p_{g_i}^t=0,  \,\ i=2,...,N\\  \label{LMP1}
&\lambda_{1}^t=...=\lambda_{N}^t=\lambda^t=\pi_1,\\  
&\hat{p}_{n_i}^t=0, \,\ 
\hat{p}_{n_{iN}}^t=0, \,\  i=2,...,N.
\end{align}

\item Case II: $\hat{p}_{n_{1N}}^t >  \hat{p}_{n_1}^t + \bar{p}$

If $\hat{p}_{n_{1N}}^t >  \hat{p}_{n_1}^t+\bar{p}$ then the congestion constraint of the line connecting bus $1$ to $2$ is an active constraint. Because of part $b)$ of Assumption~\ref{assumption3}, the generator at bus~$1$ supplies the power $\hat{p}_{n_1}^t+\bar{p}$ and the rest of the generators supply the remaining load, $p_{g_1}^t=\hat{p}_{n_1}^t+\bar{p}$. It can be deduced from part $b)$ of Assumption~\ref{assumption3} that there exists an $k$ such that $CVaR_{\alpha}^{W}(\sum_{i=k+1}^{N} (p_{d_i}^t-p_{r_i}^t)) \leq \bar{p}$
and $CVaR_{\alpha}^{W}(\sum_{i=k}^{N} (p_{d_i}^t-p_{r_i}^t)) > \bar{p}$. These inequalities imply that the $k$th generator is able to supply the total required power for bus $k$ through $N$, without violating the congestion constraint. Therefore,

\begin{align}
& p_{g_i}^t=CVaR_{\alpha}^{W}(p_{d_i}^t-p_{r_i}^t), \,\ i=2,...,k-1\\
& p_{g_k}^t=CVaR_{\alpha}^{W}(\sum_{i=K}^{N} (p_{d_i}^t-p_{r_i}^t))-\bar{p}\\
& p_{g_i}^t=0, \,\ i=k+1,...,N,\\ 
&\hat{p}_{n_1}^t=CVaR_{\alpha}^{W}(p_{d_1}^t-p_{r_1}^t),\\
&\hat{p}_{n_{1N}}^t=CVaR_{\alpha}^{W}(\sum_{j=1}^{N} (p_{d_j}^t-p_{r_j}^t)),\\
&\hat{p}_{n_i}^t=CVaR_{\alpha}^{W}(p_{d_i}^t-p_{r_i}^t)-\bar{p}, \,\ i=2,...,k\\
&\hat{p}_{n_{iN}}^t=CVaR_{\alpha}^{W}(\sum_{j=i}^{N} (p_{d_j}^t-p_{r_j}^t))-\bar{p}, \,\ i=2,...,k
\end{align}
\begin{align}
& \hat{p}_{n_i}^t=0, \,\ i=k+1,...,N\\
& \hat{p}_{n_{iN}}^t=0, \,\ i=k+1,...,N.\\ \label{LMP2}
&\lambda_i^t=\pi_i, \,\ i=1,...,k-1\\ \label{LMP3}
&\lambda_i^t=\pi_{k}, \,\ i=k,...,N.
\end{align}
\end{itemize}
Below, we formulate $\hat{p}_{n_i}^t$ and $\hat{p}_{n_{iN}}^t$ for all buses $i=1, 2,...,N$ as follows
\begin{align}\label{eq:recursive}
& \hat{p}_{n_{(i+1)}}^t(\hat{p}_{n_i}^t,\hat{p}_{n_{iN}}^t,p_{g_{i}}^t)=  \\ \notag
& \,\ \,\ \,\ \,\  \,\ \,\ \,\ \,\ \,\ \,\ \,\  \begin{cases}
0 & p_{g_i}^t=\hat{p}_{n_{iN}}^t, \\
CVaR_{\alpha}^{W}(p_{d_{i+1}}^t-p_{r_{i+1}}^t)-\bar{p} & p_{g_i}^t = \hat{p}_{n_i}^t+\bar{p},\\ 
\end{cases}
\end{align}
and $\hat{p}_{n_{(i+1)N}}^t=\hat{p}_{n_{iN}}^t-p_{g_{i}}^t$. The planned power at bus $i$ is given as
$p_{g_i}^t(\hat{p}_{n_i}^t,\hat{p}_{n_{iN}}^t)=\min (\hat{p}_{n_i}^t+\bar{p},\hat{p}_{n_{iN}}^t)$.

\begin{remark}\label{remark3}It can be concluded from (\ref{eq:recursive}) that the maximum planned power in the worst case scenario is equal to $\sum_{i=1}^{N} CVaR_{\alpha}^{W} (p_{d_i}^t-p_{r_i}^t)$. Using the subadditivity property of the $CVaR_{\alpha}^{W}$
\begin{align}\label{cong:upper}
CVaR_{\alpha}^{W}(\sum_{i=1}^{N} (p_{d_i}^t-p_{r_i}^t)) \leq \sum_{i=1}^{N} CVaR_{\alpha}^{W}  (p_{d_i}^t-p_{r_i}^t).
\end{align}
Inequality (\ref{cong:upper}), implies an upper-bound on the planned power resulting from net-load uncertainties and congestion.
\end{remark}

Remark~\ref{remark3} and inequality~(\ref{cong:upper}) imply that the uncertainties in renewable generation at a given bus can lead to a higher estimation of non-renewable generation at that bus. This leads to an unnecessary planning for the non-renewable generator. In the real-time, this leads to the dispatch of more expensive generators because of the risk of active congestion constraints in the zone of cheaper generators. Relieving the congestion constraints reduces the effect of renewable uncertainties on the economic efficiency of the dispatch strategy. This study can be extended to determine the profitability of removing the congestion at specific buses. 

\section{Non-Renewable Generator Revenue Function} \label{sec:nonren}

Loads $\{(p_{d_{1}}^{tk},...,p_{d_{N}}^{tk}) \}_{t=1}^{T}$ and renewable realizations $\{(p_{r_{1}}^{tk},...,p_{r_{N}}^{tk}) \}_{t=1}^{T}$ are deterministic functions of time for a given contingency scenario $k$. We assume that the contingency scenario $k$ may occur with probability $\psi^{k}_{W}$. We allow $k \in \{1,...,K_{W}\}$, where the number of scenarios $K_{W}$, the probability $\psi^{k}_{W}$, and renewable generation realizations depend on the installed capacity of renewables $\{w_{i}\}_{i=1}^{N}$ at all locations in the grid. Let $p_{g_{i}}^{tk}$ be the active power of the generator at bus $i$, time $t$ and scenario $k$. Actual active power generation $p_{g_{i}}^{tk}$ can be different from the corresponding committed power $p_{g_{i}}^{t}$. Let $\Delta p_{g_{i}}^{tk}=p_{g_{i}}^{t}-p_{g_{i}}^{tk}$ be the deviation from committed power $p_{g_{i}}^{t}$. Let $rp_{g_{i}}^{\max}$ be the limit on the contingency reserve of the generator at bus $i$. The power contingency reserve $rp_{g_{i}}^{t}$ of generator at bus $i$ and time $t$ must satisfy the following inequalities:
\begin{align}\label{ineq:activepowerreserve}
0 \leq \Delta p_{g_{i}}^{tk} \leq  rp_{g_{i}}^{t} \leq rp_{g_{i}}^{\max}.
\end{align}

Inequalities (\ref{ineq:activepowerreserve}) imply that the reserve quantities must be large enough to meet the demanded power in all contingency scenarios. %The upward and downward active power deviations are bounded by the physical ramp rates $\delta p_{g_{i}}^{+\max}$ and $\delta p_{g_{i}}^{-\max}$. 
The ramp reserve $\delta p_{g_{i}}^{t}$ of generator at bus $i$, for all time $t$ and scenarios $k, \acute{k} \in \{1,...,K_{W}\}$, must satisfy the following inequalities:
\begin{align} \label{ineq:activecont}
-\delta p_{g_{i}}^{\max} \leq -\delta p_{g_{i}}^{t} \leq p_{g_{i}}^{tk}-p_{g_{i}}^{(t+1)\acute{k}}  \leq \delta p_{g_{i}}^{t} \leq \delta p_{g_{i}}^{\max}.
\end{align}
Some ISOs, i.e. the Midwest ISO, provide cost recovery for their eligible committed generators through a revenue sufficiency guarantee. This cost recovery is to ensure enough energy resources are committed to the market to meet the load and reserve obligations. Because ISOs are non-profit organizations, the cost recovery payment is reflected in the market price of energy. We define $\lambda^{W}$ as the cost recovery per unit of generated power. $\lambda^{W}$ is the average of no-load cost, start-up cost, reserve cost and ramp-rate cost per unit of supplied power. $\lambda^{W}$ is affected by the penetration of the renewable energies. We consider two different scenarios in which the ISO may or may not consider the cost recovery. In the case without cost recovery the market price of energy is lower but there is a risk that the profit of non-renewable generators will decline significantly. We define $\lambda^{W}$ for the case with the cost recovery ($CR=1$) and without cost recovery ($CR=0$) as follows,

\begin{align}\label{eq:recovv} 
H=\sum_{t=1}^{T} \sum_{i=1}^{N}  {[} & \beta_{i} I_{i}^{t}+S_{i} I_{i}^{t} (1-I_{i}^{t-1})
 +f_{rp}(rp_{g_{i}}^{t})
+f_{\delta p^{}}(\delta p_{g_{i}}^{t}) {]}
\end{align}

\begin{align}\label{eq:recov}
    \lambda^{W}(CR)= 
\begin{cases}
    \frac{H}{\sum_{t=1}^{T} \sum_{i=1}^{N}  p_{g_{i}}^t},& \text{CR=1 }\\
    0,              & \text{CR=0}
\end{cases}.
\end{align}

The first term in (\ref{eq:recovv}) is the no-load cost, second term is the start-up cost, third term is the reserve cost and fourth term is the ramp cost. % and the last term is the downward ramp cost. %We assume that committed power of generator $j$th at node $i$ is equal to $p_{{g_{i}}}^{t}$, Assumption~\ref{assumption2} part $c)$. 
Let $f_{p}(p_{{g_{i}}}^{t})$ be the cost of generating power $p_{{g_{i}}}^{t}$. If the renewable generators were dispatchable, the non-renewable generators would thus expect to make a profit through this participation in the market as given by%The total profit of non-renewable generators is given as

\begin{align}\label{nonrenprofit}
R(P,CR)=\sum_{t=1}^{T} \sum_{i=1}^{N} 
\Big{[}  & p_{{g_{i}}}^{t} \big{(}\lambda_{i}^{t}-\lambda^{W}(1-CR)\big{)}
-f_{p}(p_{{g_{i}}}^{t}) \Big{]}.
\end{align}
Where $\lambda_{i}^{t}$ is the LMP at bus $i$ and time $t$, and is given by (\ref{eq:price1}), (\ref{eq:price2}) and (\ref{eq:price3}) for the case without congestion constraint. The LMPs for the case with congestion constraints (and for a radial grid) is given by (\ref{LMP1}), (\ref{LMP2}) and (\ref{LMP3}). The recovery cost $\lambda^W(CR)$ is given by (\ref{eq:recov}). The first term in (\ref{nonrenprofit}) is the revenue, the second term is the recovery cost, and the third term is the cost of generating $p_{g_{i}}^{tk}$. % the fourth term is the deviation cost from $p_{g_{ij}}^{tc}$. 
However, due to the unpredictable renewable generation, the non-renewable generators will not make this profit. The revenue they earn is determined by the power they actually sell, which in turn, is a function of the renewable realization. We now calculate the profit that the non-renewable generator actually makes. 

\begin{align}\label{nonrenprofitreal}
\tilde{R}(P^{K},CR)=\sum_{t=1}^{T} \sum_{i=1}^{N}  \sum_{k=1}^{K_{W}}  \psi^{k}_{W} \Big{[}  & p_{{g_{i}}}^{tk} \big{(}\lambda_{i}^{t}-\lambda^{W}(1-CR)\big{)}\\ \notag
&-f_{p}(p_{{g_{i}}}^{tk})  \Big{]}.
\end{align}

Where $p_{{g_{i}}}^{tk}$ is the dispatched power of generator at node $i$ given the load and renewable realization at scenario $k$. If the renewable generation is larger than the load then the renewables are paid based on the committed price of energy ($\lambda_{i}^{t}$), proportional to the load and excess renewable generation is curtailed. If the realized renewables is smaller than the load, then the renewable generator is paid based on the committed price of energy for all the generations. 

Both price $\lambda_{i}^{t}$ and cost recovery $\lambda^W(CR)$ are affected by the penetration level of the renewables. The higher intermittency and uncertainty of renewable generations increases the number of  committed generators and subsequently $\lambda^{W}$. Higher penetration of the renewables may lower the locational marginal prices $\lambda_{i}^{t}$, because of decreasing the expected net-load for the non-renewables (this argument is only valid if the uncertainty of renwables does not rise significantly).  Higher penetration of the renewables will increase the deviation power $\Delta p_{g_{i}}^{tk}$.

We also calculate the loss in profit of the non-renewable generators due to the fact that the renewable generators could not be predictably dispatched and yet the grid absorbed the renewable supply when available and expected the non-renewables to compensate for the intermittence of the renewables. This defines the cost of deviation suffered by the non-renewables as
\begin{equation}\label{eq:devcost}
\text{Deviation Cost} = R(P,CR)-\tilde{R}(P^{K},CR).
\end{equation}

 The deviation cost in (\ref{eq:devcost}) is the difference between the profit of the non-renewable generator, if the committed power is dispatched, minus the expected profit over the contingency scenarios. Decreasing the profit in higher penetration of the renewables disincentives non-renewables to remain in the market. The profit of non-renewable generators must be big enough that they at least break even over the levelized cost of energy through the life-time of the non-renewable generators. This problem become more crucial when the renewables have zero marginal cost and non-renewables are losing the market to the non-renewables. This may cause non-renewables to offer higher prices of energy to compensate their levelized cost of energy at times of renewable shortages and consequently more fluctuation in the market price of energy.% This problem arises in a market where renewables and non-renewables are separate entities.   

\section{Numerical Analysis}\label{sec:IV}

In this section, we study the effect of increasing the penetration of the renewables energy and desired level of the reliability on the market price of energy and profit of the non-renewables. We consider a grid of seven generators with physical constraints and cost parameters in Table~\ref{table:genprm}. The generators cost parameters and physical constraints are provided based on \cite{Frank}.

\begin{table}[h!]
\caption{The generators cost parameters and physical constraints \cite{Frank}.}

\label{table:genprm}
\begin{center}
\begin{tabular}{cccccccc}
\hline
Generator & 1 & 2 & 3 & 4 \\ [1em]
Production Cost ($\$ /MWh$) & 7.37 & 22.23 & 31.55 & 176.05 \\ [1em]
%Min. ($MW$) & 0 & 0 & 0 & 0 \\ [1em]
Max. ($MW$) & 400 & 155 & 76 & 197 \\ [1em]
Hot Start Cost ($\$$) & 0 & 2258.6 & 1412.5 & 14182.5 \\ [1em]
Cold Start Cost ($\$$) & 0 & 616.2 & 1412.5 & 8106.9 \\ [1em]
Ramp Up and Down ($MW/h$) & 400 & 155 & 76 & 197 \\ [1em]
\hline
\hline
Generator & 5 & 6 & 7 \\ [1em]
Production Cost ($\$ /MWh$) & 180.75 & 241.91 & 315.81\\ [1em]
%Min. ($MW$) & 0 & 0 & 0\\ [1em]
Max. ($MW$) & 100 & 12 & 20\\ [1em]
Hot Start Cost ($\$$) & 10357.8 & 1244.4 & 109.5\\ [1em]
Cold Start Cost ($\$$) & 4575.0 & 695.4 & 109.5\\ [1em]
Ramp Up and Down ($MW/h$) & 100 & 12 & 20\\ [1em]
\hline
\end{tabular}
\end{center}
\end{table}
\begin{figure}[!htbp]
\centering
\subfigure[Committed power versus the reliability level.]{
\includegraphics[scale=0.4]{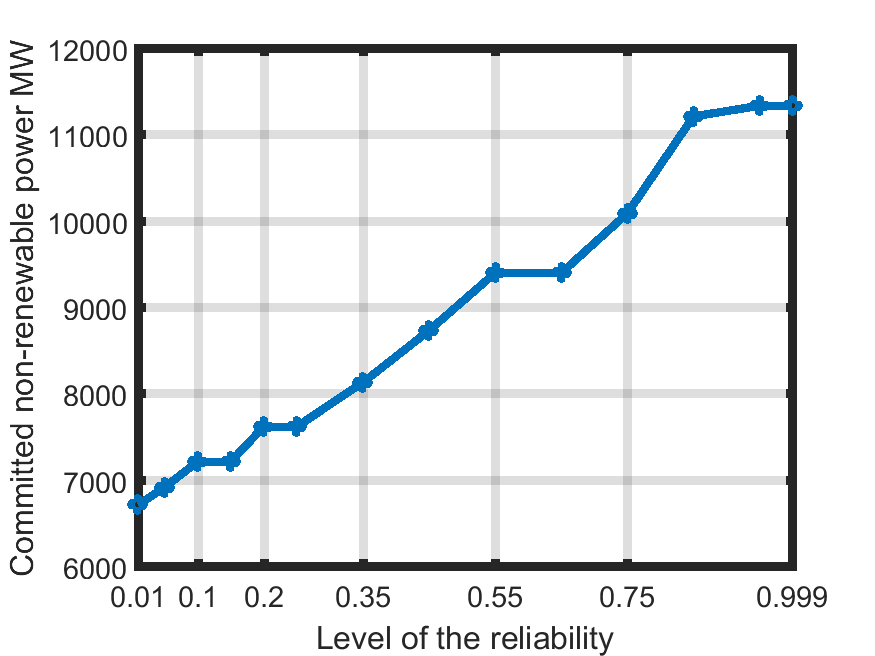}
\label{fig:dayaheadpwrreli}
}
\subfigure[Price of energy for committed energy versus the reliability level.]{
\includegraphics[scale=0.4]{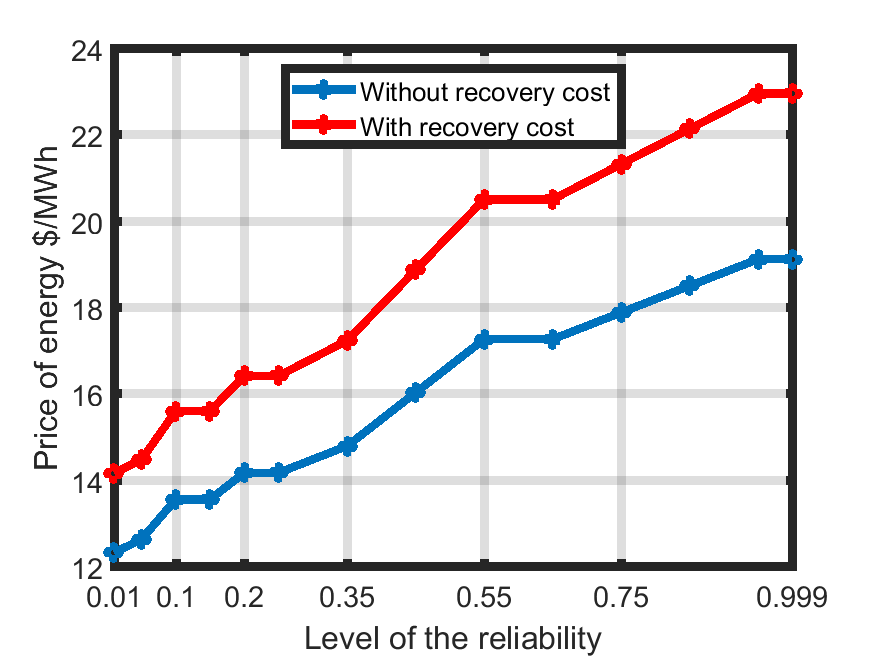}
\label{fig:dayaheadpricereli}
}
\subfigure[Expected profit of the non-renewable generator versus the reliability level.]{
 \includegraphics[scale=0.4]{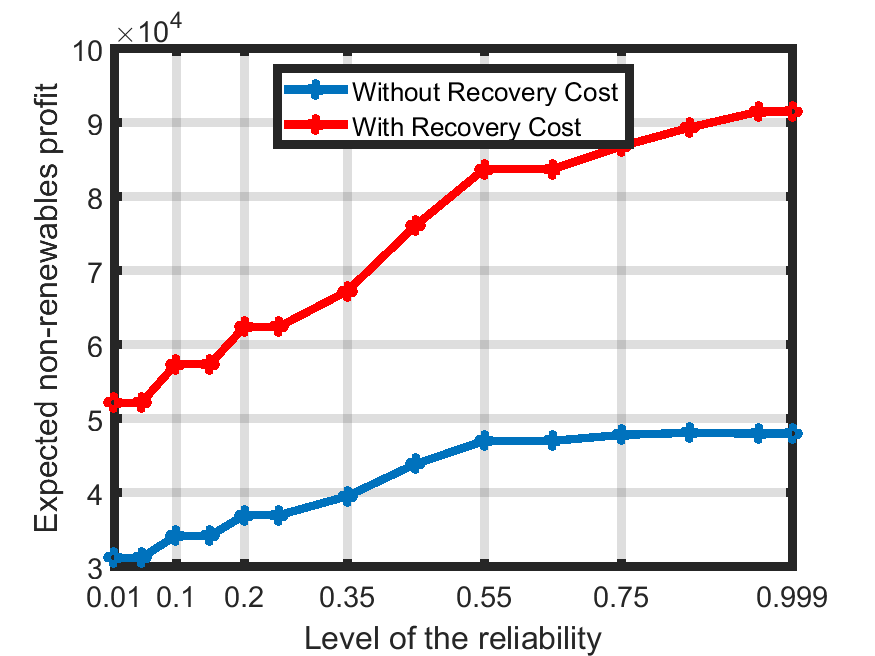}
 \label{fig:expecprofitreli}
}
\end{figure}

\begin{figure}[!htbp]
\centering
\subfigure[Recovery cost versus the reliability level.]{
 \includegraphics[scale=0.4]{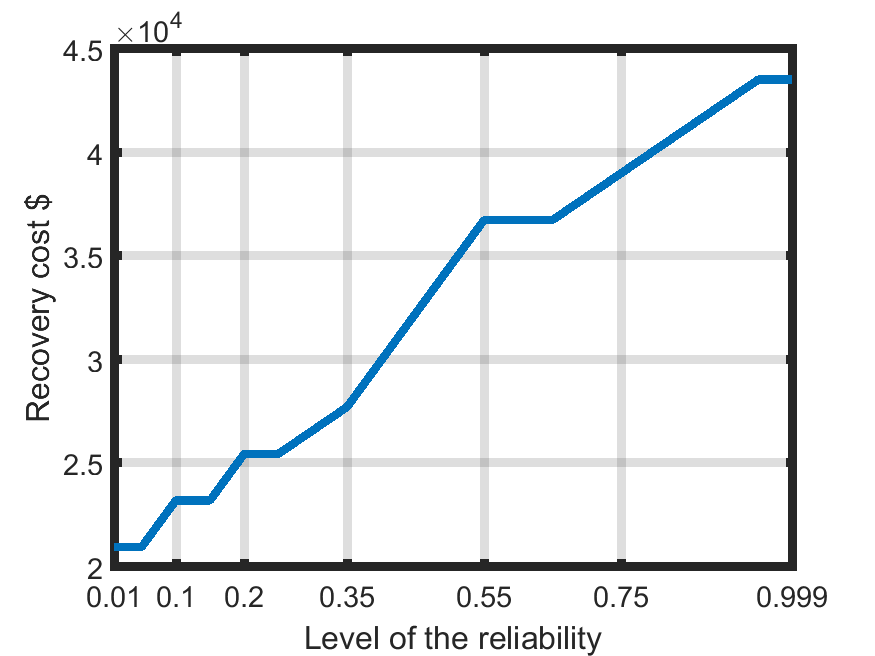}
 \label{fig:expeccompcost}
}
%\subfigure[Expected price of energy versus the reliability level.]{
% \includegraphics[scale=0.4]{exp_realtime_price_reliability.png}
% \label{fig:expectrealtimepricereli}
%}
% \subfigure[Expected profit of the renewable generator.]{
% \includegraphics[scale=0.5]{ren_profit_reliability.png}
% \label{fig:exprenewprftreli}
% }
\caption{Reliability level and non-renewable energy production under simulation scenarios}
\label{fig4}
\end{figure}
We first fix the penetration level of the renewables at $0.9$ percent and study how the market price of energy, generators profit and planned power varies for different desired level of reliability. In the Figure~\ref{fig:dayaheadpwrreli}, and \ref{fig:dayaheadpricereli} the committed power and price versus the reliability level is shown respectively. The committed power and price are increasing in the desired level of reliability. 

The expected profit of the non-renewables with and without recovery cost versus the reliability level is shown in Figure~\ref{fig:expecprofitreli}. In Figure~\ref{fig:expeccompcost}, it is shown that the recovery cost is increasing in the reliability level. Higher reliability level leads to a higher planning for the non-renewables and increase the recovery cost.

%The expected market price of energy with the recovery cost and without the recovery cost is shown in the Figure ~\ref{fig:expectrealtimepricereli}. Higher reliability level causes more planning for committed power in the day-ahead market, consequently the expensive peaker generators are less used in the contingency scenarios therefore the expected price decreases. 

\begin{figure}[!htbp]
\centering
\subfigure[Committed power versus the penetration level.]{
\includegraphics[scale=0.4]{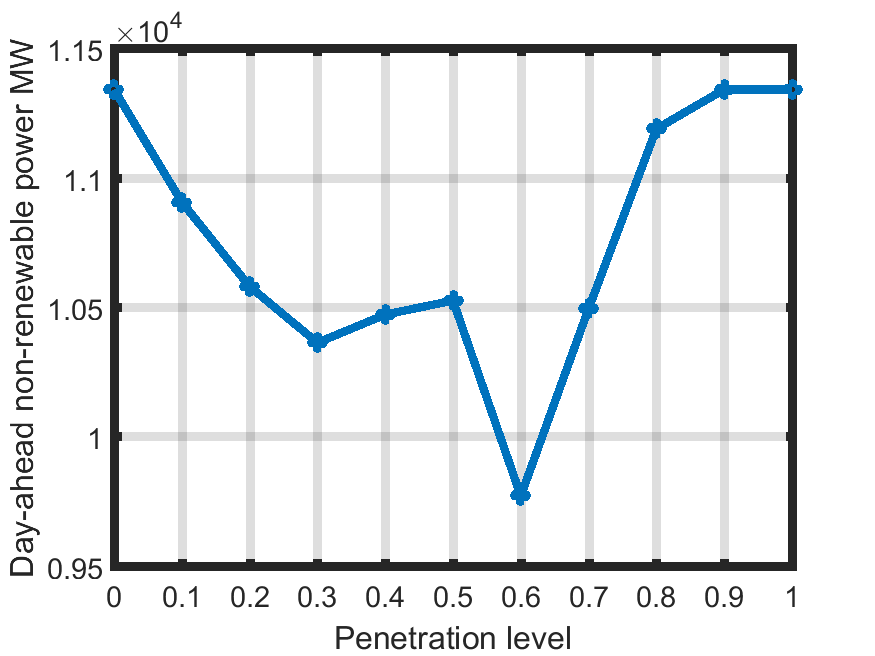}
\label{fig:dayaheadpowerpen}
}
\subfigure[Price of energy for committed energy versus the penetration level.]{
\includegraphics[scale=0.4]{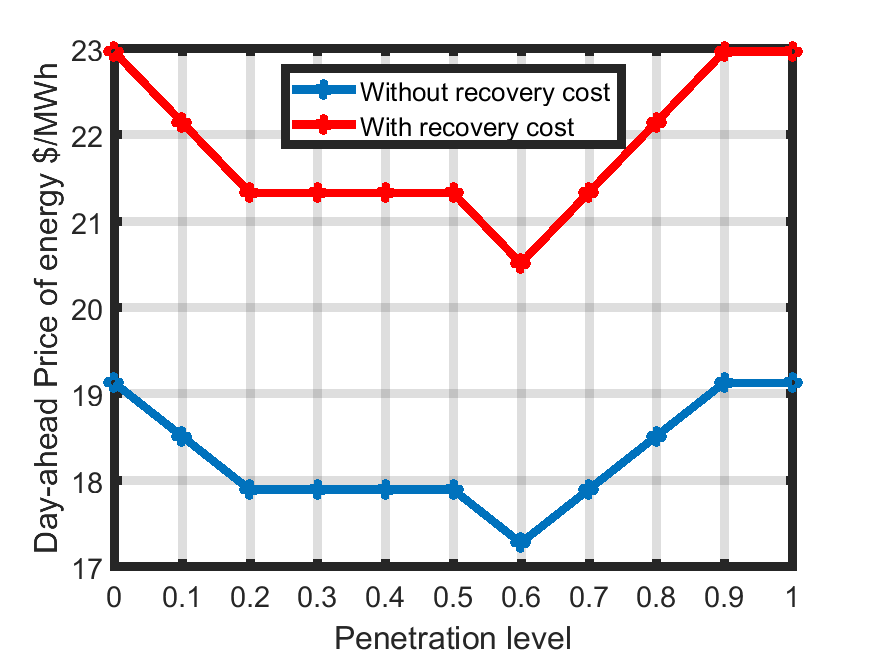}
\label{fig:dayaheadpricepen}
}
% \subfigure[Day-ahead profit versus the penetration level.]{
% \includegraphics[scale=0.5]{dayaheadprofit_penet.png}
% \label{fig:contingency}
% }
\subfigure[Deviation cost versus the penetration level.]{
\includegraphics[scale=0.4]{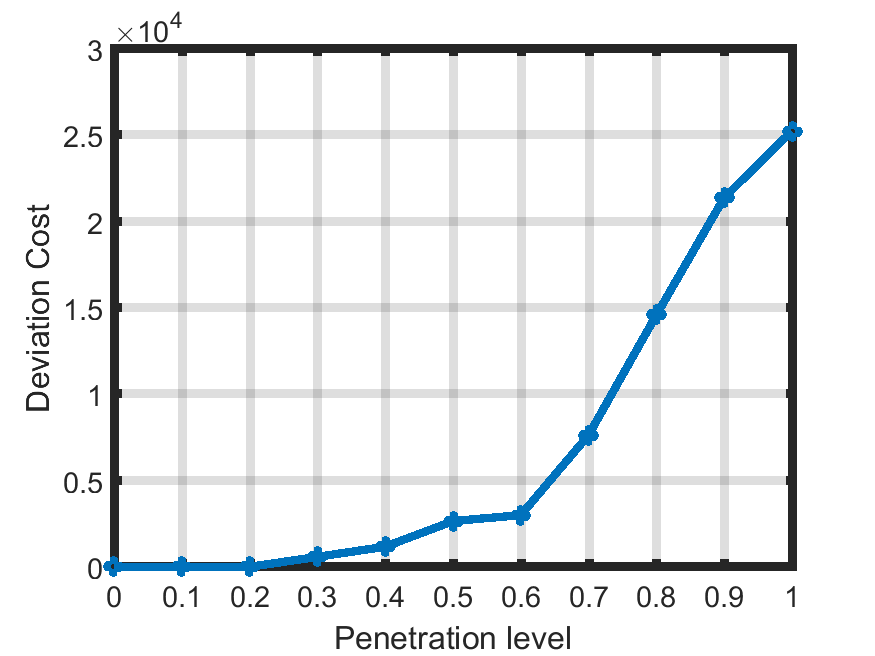}
\label{fig:deviationcost}
}
\end{figure}

\begin{figure}[!htbp]
\centering
\subfigure[Expected profit of the non-renewable generator.]{
\includegraphics[scale=0.4]{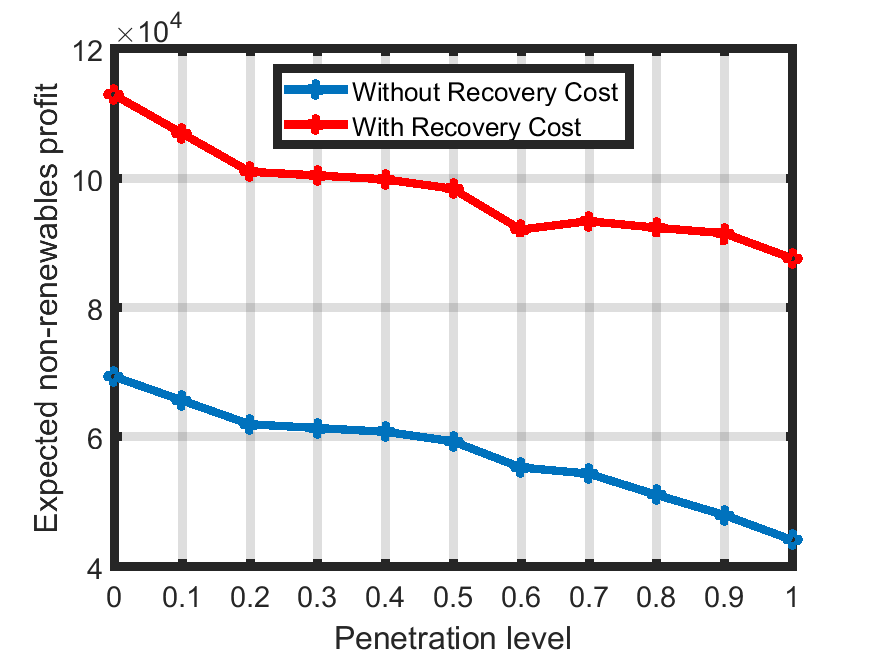}
\label{fig:nonrencompsprofitpen}
}
\subfigure[Expected compensation cost versus the penetration level.]{
\includegraphics[scale=0.4]{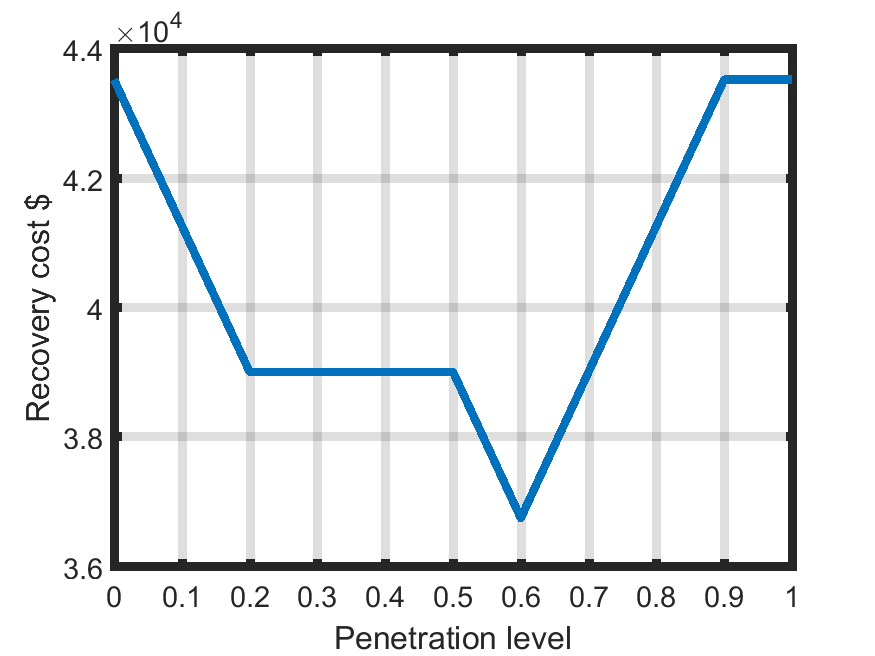}
\label{fig:recoverycostpen}
}
\subfigure[Expected profit of the renewable generator.]{
\includegraphics[scale=0.4]{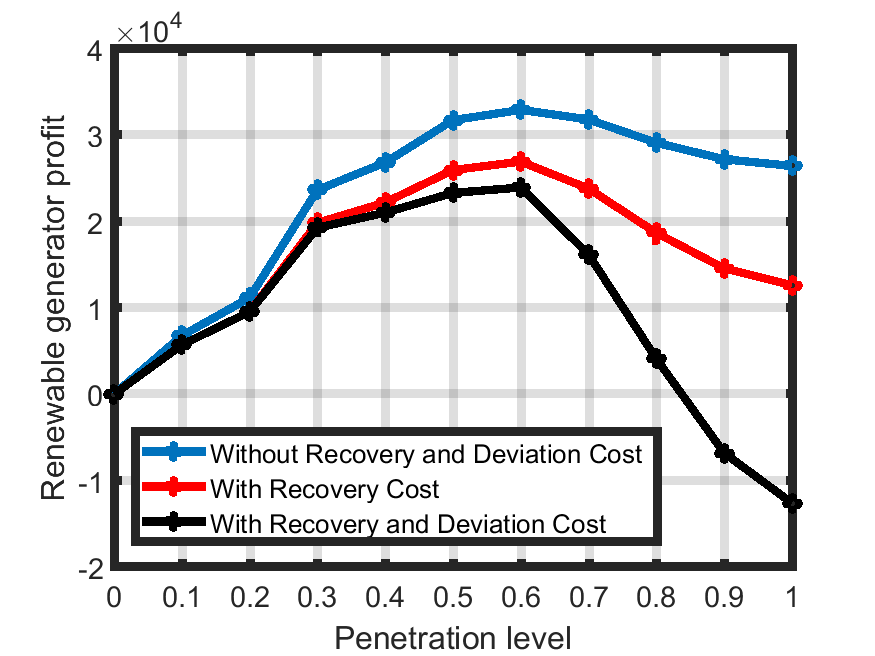}
\label{fig:expectedrenprofitpen}
}
\caption{Renewables penetration level and non-renewable energy production under simulation scenarios}
\label{fig5}
\end{figure}

Below, we fix the desired level of reliability at $0.9$. We vary the penetration of the renewables and study how the market price of energy, generator profits and planned power vary for different levels of penetration. Figures \ref{fig5}(a) %~\ref{fig:dayaheadpowerpen}
, and~\ref{fig:dayaheadpricepen} show the committed power, and price versus renewables penetration level respectively. The total amount of dispatched non-renewable energy first decreases as the penetration of the renewables increases because the net-load decreases. After reaching a threshold, the required power from non-renewables increases due to uncertainty about the production of renewables. Figure~\ref{fig:dayaheadpricepen} shows that there is an intermediate point in the capacity of renewables that the market clearing price of energy has reached its minimum. The low marginal cost of renewables is traded-off with the additional non-renewable generators dispatched to counter the increased risk from a higher penetration of renewables.

The expected deviation cost is shown in Figure~\ref{fig:deviationcost}. It is shown that the deviation cost is increasing in the penetration level. It rises rapidly if the penetration increases more than $60$ percent. The expected profit of the non-renewables with and without recovery cost versus the penetration level is shown in Figure~\ref{fig:nonrencompsprofitpen}. It is shown that the profit of non-renewables is decreasing in the penetration level of the renewables. In Figure~\ref{fig:recoverycostpen}, it is shown that the recovery cost has a behaviour similar to the market price of energy. A higher penetration level leads to a higher uncertainty for the renewable generations and increases the recovery cost. It is shown in Figure~\ref{fig:recoverycostpen} that the recovery cost rises if the penetration increases more than $60$ percent. Therefore the price of energy increases after $60$ percent, Figure~\ref{fig:dayaheadpricepen}. In Figure~\ref{fig:expectedrenprofitpen}, the expected profit of renewables with and without the recovery and deviation cost is shown. It is observed that the profit of renewables is decreasing in the penetration level, if it is more than $60$ percent.

\section{Conclusion}\label{sec:V}
%\blindtext

We obtain an analytical formula to quantify the amount of committed power and market clearing price of energy as a function of the desired level of reliability and penetration level of the renewable energies. We also obtain an analytical upper-bound on the amount of committed power resulting from net-load uncertainties and congestion constraints. It is observed that the power and price of energy decrease in renewables penetration level until a threshold then start rising due to uncertainties in the renewables realizations. It is observed that the amount of committed power and the market price of energy are increasing in the reliability level. The market price of energy may increase due to the uncertainty of the net-load. The expectation that increasing the penetration of renewable energy reduces the market price of energy would be untrue if uncertainty in renewable energy generation is increased by a higher penetration level. The recovery cost of non-renewables generators, to compensate start-up costs, no-load cost, ramp reserve cost and contingency reserve cost, increases in the reliability level. The deviation of the committed power from dispatched power is increasing in the penetration level of the renewables.  %\rev{reference  \cite{ghavidel2, re}}

%Developing strategies to allocate the congestion costs in the presence of high penetration of renewables may be another direction for future studies. In conclusion, with higher penetration of renewables, investment costs for removing congestion must be more effectively incorporated into the market operation status. An alternative study can be directed toward developing an optimal strategy for non-renewables to invest in renewable energy generation to avoid a scenario where renewables have a higher share of the market. A game theoretical framework can be developed to find the equilibrium penetration level of renewables for the market price of energy in which the non-renewable, renewables and consumers agree on an equilibrium strategy. Another study could focus on ways to mitigate the deviation of the spot market from the day-ahead market in the presence of renewables. This problem can be more complex in a market where the utilities are looking to take advantage of the differences between the day ahead and real-time market price of energy.

\section{Appendix}\label{sec:VI}
\textbf{Proof of Theorem~\ref{thm1}:}

%Higher uncertainty in the renewable energy generation and load model increases the likelihood that the congestion constraints (\ref{ineq:cons}) will become active constraints in more number of lines. Subsequently, generators with higher marginal prices are dispatched. For this reason, the locational marginal prices (LMPs)  increase at some and decrease at other buses.

%Therefore, the optimization problem (\ref{optim}) is disjoint in time. 
The Lagrangian function for the constrained optimization problem (\ref{optim}) for the given time $t$ can be written as:
\begin{align}\label{ISO}
L^t(&P^t)= \underset{P^t}{\min}   \Bigg{[}  \sum_{i=1}^{N}  \big{[} \pi_{i} p_{g_{i}}^{t}+\pi_{r} p_{r_i}^{t} \big{]} \\ \notag
&+  \sum_{i=1}^{N} \lambda_i^{t} \Big{[}  - p_{g_{i}}^{t}+p_{d_i}^{t}-p_{r_i}^{t}+\sum_{j \in C_{i}} b_{ij}  (\theta_i^t-\theta_j^t)  \Big{]} \\ \notag
&+\sum_{i=1}^{N} \sum_{j \in C_{i}} \tilde{\mu}_{ij}^{t} \big{[}  b_{ij} (\theta_i^t-\theta_j^t)-{p}_{ij}^{\max} \big{]}\\ \notag
%&+\mu_{21} \Big{[}  Y_{21} (\theta_2^t-\theta_1^t)-\bar{p}_{21} \Big{]}\\ \notag
&+\sum_{i=1}^{N} \mu_{i}^t (p_{g_{i}}^{t}-p_{g_{i}}^{\max})
+\sum_{i=1}^{N} \bar{\mu}_{i}^{t} (p_{g_{i}}^{\min}-p_{g_{i}}^{t}) \Bigg{]}.
\end{align}
The optimal solutions for $p_{g_{i}}^{t}$ and $\theta_i^t$ are obtained by solving the Karush-Kuhn-Tucker (KKT) conditions (\ref{opt1})-(\ref{nonneg}) for all $i=1,...,N$ as follows.
\begin{align}\label{opt1}
&\pi_{i}-\lambda_i^t+\mu_{i}^{t}-{\bar{\mu}}_{i}^{t}=0, \\ \label{opt2}
&\sum_{j \in C_i} b_{ij} [ \tilde{\mu}_{ij}^t-{\tilde{\mu}}_{ji}^{t}+\lambda_i^t-\lambda_j^t]=0, \\ \label{feas1}
&p_{r_i}^{t}-p_{d_i}^{t}+ p_{g_{i}}^{t}=\sum_{j \in C_{i}} b_{ij} (\theta_i^t-\theta_j^t),\\ \label{feas2}
%\end{align}
%\begin{align} \label{feas2}
&b_{ij} (\theta_i^t-\theta_j^t) \leq {p}_{ij}^{\max},\\ \label{feas3}
& p_{g_{i}}^{\min} \leq p_{g_{i}}^{t} \leq p_{g_{i}}^{\max}\\ \label{cs1}
&\tilde{\mu}_{ij}^{t} \big{(}b_{ij} (\theta_i^t-\theta_j^t) - {p}_{ij}^{\max}\big{)}=0 \\ \label{cs3}
&\mu_{i}^{t} (p_{g_{i}}^{t}-p_{g_{i}}^{\max})=0, \\ \label{cs2}
&\bar{\mu}_{i}^{t} (p_{g_{i}}^{\min}-p_{g_{i}}^{t})=0, \\ \label{nonneg}
&\mu_{i}^{t} \geq 0, \bar{\mu}_{i}^{t} \geq 0.
\end{align}
The equations (\ref{opt1}) and (\ref{opt2}) are obtained by solving $\frac{\partial L}{\partial p_{g_{i}}^{t}}=0$ and $\frac{\partial L}{\partial \theta_{i}^t}=0$ respectively. (\ref{feas1})-(\ref{feas3}) are the feasibility conditions. (\ref{cs1})-(\ref{cs2}) are the complementary slackness conditions. The inequalities (\ref{nonneg}) are the non-negativity conditions.

\textbf{Proof of Theorem~\ref{thm2}:} From \cite{Rockafellar}
\begin{equation}\label{eq:cvar00}
CVaR_{\alpha}^{W}(n^t)=\min_{\eta \in \Re} \Big{\{} \eta + \frac{1}{1-\alpha} E [n^t-\eta ]^+ \Big{\}}.
\end{equation}
We first prove that the minimizer of (\ref{eq:cvar00}) is non-negative. Suppose the minimizer of (\ref{eq:cvar00}) is negative ($\eta \leq 0$). Let $\eta=-\eta^{+}$, where $\eta^{+}=|\eta|$, then
\begin{align}\label{eq:cvar0000}
&\arg\min_{\eta \in \Re} \Big{\{} \eta + \frac{1}{1-\alpha} E [n^t-\eta ]^+ \Big{\}} \\ \notag
%&=\arg\min_{\eta^{+} \in \Re^+} \Big{\{} -\eta^{+}+\frac{1}{1-\alpha} E [n^t+ \eta^{+}] \Big{\}}\\ \notag
&=\arg\min_{\eta^{+} \in \Re^+} \Big{\{} \frac{\alpha \eta^{+}}{1-\alpha}+\frac{1}{1-\alpha} E [n^t] \Big{\}}.
\end{align}
The equality can be explained by $n^t \geq 0$, and the linear property of the expectation. It is evident that the minimizer of (\ref{eq:cvar0000}) is $\eta^{+}=0$. Therefore, the minimizer of (\ref{eq:cvar00}) is non-negative ($\eta \geq 0$). Because of the convexity of $CVaR_{\alpha}^{W}$, the minimizer of (\ref{eq:cvar00}) is obtained by taking the derivative of $\eta + \frac{1}{1-\alpha} E [n^t-\eta ]^+$ with respect to $\eta$ as
\begin{equation}
\eta^*= VaR_{\alpha}^{W}(\sum_{i=1}^{N} s_{i}^t)-\sum_{i=1}^{N}p_{g_{i}}^t.
\end{equation}
By substituting $\eta^*$ in $\eta + \frac{1}{1-\alpha} E [n^t-\eta ]^+$
\begin{align}\label{eq:fcvarr}
CVaR_{\alpha}^{W}(n^t)=-\sum_{i=1}^{N}p_{g_{i}}^t+CVaR_{\alpha}^{W}(\sum_{i=1}^{N} s_{i}^t).
\end{align}

\textbf{Proof of Theorem~\ref{thm3}:}
We assume that the generator $i$ is able to raise its generation from $p_{g_{i}}^{\min}$ to $p_{g_{i}}^{\max}$. Therefore, the optimization problem (\ref{optimm}) is disjoint in time. The Lagrangian function for the constrained optimization problem (\ref{optimm}), at time $t$, can be written as:
\begin{align}\label{ISO}
\underset{P^t}{\min} \,\ L(P^t)=\underset{P^t}{\min} \underset{P_{D},P_{r}}{E} & \sum_{i=1}^{N} \Big{[}  \pi_{i} p_{g_{i}}^{t}+\pi_{r_i} p_{r_i}^{t}  + \mu_{i}^t (p_{g_{i}}^{t}-p_{g_{i}}^{\max}) \\ \notag &\,\ \,\ \,\ \,\ \,\ +\bar{\mu}_{i}^t (p_{g_{i}}^{\min}-p_{g_{i}}^{t}) \Big{]}\\ \notag
&+ \lambda^{t} \big{(}  -\sum_{i=1}^{N} p_{g_{i}}^{t}+CVaR_{\alpha}^{W}(\sum_{i=1}^{N} s_i^{t}) \big{)}. 
\end{align}
It is evident that the Lagrange function (\ref{ISO}) is convex in $p_{g_{i}}^{t}$. The optimal solution $P^t=(p_{g_{1}}^{t},...,p_{g_{N}}^{t})$ are obtained by solving the Karush-Kuhn-Tucker (KKT) conditions (\ref{opt100})-(\ref{nonneg100}) for all $i=1,...,N$ as follows,
\begin{align}\label{opt100}
&\pi_{i}-\lambda^t+\mu_{i}^t-\bar{\mu}_{i}^t=0,\\
& \sum_{i=1}^{N} p_{g_{i}}^{t}=CVaR_{\alpha}^{W}(\sum_{i=1}^{N} s_i^{t}),\\
& p_{g_{i}}^{\min} \leq p_{g_{i}}^{t} \leq p_{g_{i}}^{\max},\\
&\mu_{i}^t (p_{g_{i}}^{t}-p_{g_{i}}^{\max})=0, \\
&\bar{\mu}_{i}^t (p_{g_{i}}^{\min}-p_{g_{i}}^{t})=0, \\ \label{nonneg100}
&\mu_{i}^t \geq 0, \bar{\mu}_{i}^t \geq 0.
\end{align}

% Let $p_{\mathfrak{nr}}^t=(p_1^t,...,p_N^t)$. We define $(\tilde{\mu}_i,\mu_i)$ and $\lambda^t$ as the lagrange multipliers corresponding to (\ref{eq:p11})
% and (\ref{eq:twentyyya}). The lagrange function for the ISO problem is given as
% \begin{align}\label{eq:lagrange}
% L(p^t)=&E_{P_{D},P_{\mathfrak{r}}} \Big{[} \sum_{i=1}^{N}  \pi_{i} p_{i}^t+ \pi_{\mathfrak{r}} p_{\mathfrak{r}}^t \Big{]}+\lambda^t \Big{[} CVaR_{\alpha}(n^t) \Big{]} \\ \notag
% &+\sum_{i=1}^{N} \mu_i \Big{[} p_i^t-p_i^{\max}  \Big{]} +\sum_{i=1}^{N} \tilde{\mu}_i \Big{[} p_i^{\min}-p_i^t  \Big{]}.
% \end{align}
% \\

% By substituting (\ref{eq:cvar0}) in (\ref{eq:lagrange})
% \begin{align}
% L(p^t)=
% &E_{P_{D},P_{\mathfrak{r}}} \Big{[} \sum_{i=1}^{N}  \pi_{i} p_{i}^t+ \pi_{\mathfrak{r}} p_{\mathfrak{r}}^t \Big{]}\\ \notag
% &+\lambda^t \Big{[} r_1 (\sum_{i=1}^{N}p_i^t)^2-\sum_{i=1}^{N}p_{i}^t+CVaR_{\alpha}(s^t) \Big{]}\\ \notag
% &+\sum_{i=1}^{N} \mu_i \Big{[} p_i^t-p_i^{\max}  \Big{]} +\sum_{i=1}^{N} \tilde{\mu}_i \Big{[} p_i^{\min}-p_i^t  \Big{]}.
% \end{align}
% The necessary Karush-Kuhn-Tucker (KKT) conditions for the ISO's problem are
% \begin{align}\label{eq:KKT1}
% &\pi_i+2 \lambda^t r_1 \sum_{i=1}^{N} p_i^t-\lambda^t+\mu_i-\tilde{\mu}_i=0,\\
% &r_1 (\sum_{i=1}^{N}p_i^t)^2+CVaR_{\alpha}(s^t)=\sum_{i=1}^{N}p_{i}^t,\\
% &p_i^{\min} \leq p_i^t \leq p_i^{\max},\label{eq:rang} \\
% &\mu_i (p_i^t-p_i^{\max})=0,\label{eq:ortho}\\
% &\tilde{\mu}_i (p_i^{\min}-p_i^t)=0,\\
% &\mu_i \geq 0, \tilde{\mu}_i \geq 0. \label{eq:KKT5}
% \end{align}

By solving (\ref{opt100})-(\ref{nonneg100}), the values of $\{\mu_i^t\}_{i=1}^{N}$, $\{\bar{\mu}_i^t \}_{i=1}^{N}$
and $\lambda^t$ are given below based on the value of $CVaR_{\alpha}^{W}(\sum_{i=1}^{N} s_{i}^t)-\sum_{i=1}^{k-1} p_{g_{i}}^{\max}$.

\begin{itemize}
\item $p_{g_k}^{\min} \leq CVaR_{\alpha}^{W}(\sum_{i=1}^{N} s_{i}^t)-\sum_{i=1}^{k-1} p_{g_{i}}^{\max} \leq p_{g_k}^{\max}$
\begin{align}
&\mu_i^t >0, \bar{\mu}_i^t=0, p_{g_i}^t=p_{g_i}^{\max} \text{ for  \,\ } i=1,...,k-1,\\
&\mu_k^t=0, \bar{\mu}_k^t=0, p_{g_k}^t=CVaR_{\alpha}^{W}(\sum_{i=1}^{N} s_{i}^t)-\sum_{i=1}^{k-1} p_{g_{i}}^{\max}, \\
&\mu_i^t=0, \bar{\mu}_i^t=0, p_{g_i}^t=0, \text{ for all } i=k+1,...,N,\label{eq:SOKKT5}\\
&\lambda^t=\pi_i+\mu_i^t-\bar{\mu}_i^t, \,\ \text{for all} \,\ i=1,...,k\\
&\lambda^t={\pi_k}.  \label{eq:result0}
\end{align}

\item $ 0 < CVaR_{\alpha}^{W}(\sum_{i=1}^{N} s_{i}^t)-\sum_{i=1}^{k-1} p_{g_{i}}^{\max} < p_{g_k}^{\min}$

\begin{align}
&\mu_i^t >0, \bar{\mu}_i^t=0, p_{g_i}^t=p_{g_i}^{\max} \,\ \text{for all}\,\ i=1,...,k-2,\\
& \mu_{k-1}^t =0, \bar{\mu}_{k-1}^t=0, \\ \notag
& p_{g_{k-1}}^t=CVaR_{\alpha}^{W}(\sum_{i=1}^{N} s_{i}^t)-\sum_{i=1}^{k-2}p_{g_i}^{\max}-p_{g_k}^{\min}, \\
&\mu_k^t=0, \bar{\mu}_k^t>0, p_{g_k}^t=p_{g_k}^{\min}, \\
&\mu_i^t=0, \bar{\mu}_i^t=0, \\
&p_{g_i}^t=0, \text{ for all } i=k+1,...,N,\label{eq:SOKKT5}\\
&\lambda^t=\pi_i+\mu_i^t-\bar{\mu}_i^t, \,\ \text{for all} \,\ i=1,...,k\\
&\lambda^t=\pi_{k-1}.  \label{eq:result}
\end{align}

\item $0< CVaR_{\alpha}^{W}(\sum_{i=1}^{N} s_{i}^t) <p_{g_1}^{\min}$

From part b) of Assumption~\ref{assumption3}, there exists an $1 \leq k \leq N$ such that 
$p_{g_{k}}^{\min} \leq CVaR_{\alpha}^{W}(\sum_{i=1}^{N} s_{i}^t) <p_{g_{k}}^{\max}$. Let $\bar{k}$ be the smallest $k$ that satisfies this condition, then
\begin{align}
&\mu_i^t=0, \bar{\mu}_i^t=0, p_{g_i}^t=0, \text{ for all } i \neq \bar{k},\\
&\mu_{\bar{k}}^t=0, \bar{\mu}_{\bar{k}}^t=0, p_{g_{\bar{k}}}^t=CVaR_{\alpha}^{W}(\sum_{i=1}^{N} s_{i}^t), \\
&\lambda^t=\pi_{\bar{k}} \label{eq:resultt}.
\end{align}
\end{itemize}

\textbf{Proof of Lemma~$1$:} The right side inequality in (\ref{neq:lem1}) is obvious. Below the left side inequality is proven.
Because of part $b)$ of Assumption~$3$,
%\begin{align}
%-(p_{k-1}^{\max}-p_{k-1}^{\min}) <- p_k^{\min}. \label{ineq:2}
%\end{align}
%By adding $p_{k-1}^{\max}$ to the both sides of inequality (\ref{ineq:2})
\begin{align}
p_{g_{k-1}}^{\min} <p_{g_{k-1}}^{\max}- p_{g_k}^{\min}. \label{ineq:3}
\end{align}
Because of $0 < CVaR_{\alpha}^{W}(\sum_{i=1}^{N} s_{i}^t)-\sum_{i=1}^{k-1} p_{g_{i}}^{\max}$
\begin{align}
p_{g_{k-1}}^{\max} < CVaR_{\alpha}^{W}(\sum_{i=1}^{N} s_{i}^t)-\sum_{i=1}^{k-2} p_{g_{i}}^{\max}. \label{ineq:4}
\end{align}
It is concluded from (\ref{ineq:3}) and (\ref{ineq:4})
\begin{align}
p_{g_{k-1}}^{\min} & < p_{g_{k-1}}^{\max}- p_{g_k}^{\min}
\\ \notag & < CVaR_{\alpha}^{W}(\sum_{i=1}^{N} s_{i}^t)-\sum_{i=1}^{k-2} p_{g_{i}}^{\max}- p_{g_k}^{\min}.
\end{align}

%\section*{Acknowledgment}
% This work was supported by the Center for Sustainable Energy at Notre Dame and NSF grant ECCS 1550016.

\bibliographystyle{IEEEtranN}
\bibliography{reference}
\begin{IEEEbiographynophoto}
{Ashkan Zeinalzadeh} received his B.S. in Electrical Engineering at the University of Shiraz, and his M.S. in mathematics and PhD in Electrical Engineering at the University of Hawaii at Manoa. He is currently with the University of Notre Dame as a postdoctoral fellow. His research interests are in the area of game theory, distributed optimization, power systems and energy markets.
\end{IEEEbiographynophoto}
\begin{IEEEbiographynophoto}
{Donya Ghavidel}
received her B.S. degree in Electrical Engineering at Iran University of Science and Technology, Tehran, Iran, 2014, and her M.S. degree in Electrical Engineering from the University of Notre Dame, Notre Dame, IN, USA, in 2016, where she is currently pursuing the Ph.D. degree in Electrical Engineering.
Her research interests lie in game theory, mechanism design, and their applications to power systems and networks.
\end{IEEEbiographynophoto}
\begin{IEEEbiographynophoto}{Vijay Gupta}
Vijay Gupta is a Professor in the Department of Electrical Engineering at the University of Notre Dame, having joined the faculty in January 2008. He received his B. Tech degree at Indian Institute of Technology, Delhi, and his M.S. and Ph.D. at California Institute of Technology, all in Electrical Engineering. Prior to joining Notre Dame, he also served as a research associate in the Institute for Systems Research at the University of Maryland, College Park. He received the 2013 Donald P. Eckman Award from the American Automatic Control Council and a 2009 National Science Foundation (NSF) CAREER Award. His research and teaching interests are broadly at the interface of communication, control, distributed computation, and human decision making.
\end{IEEEbiographynophoto}
\end{document}